\documentclass[aps,prb,preprint,floats,showpacs]{revtex4}
\usepackage{epsfig,amssymb,amsfonts,amsmath}
\usepackage{graphicx}
\input{epsf}
\begin{document}

\title{Gapless surface states in a lattice of coupled cavities: a photonic analog of topological crystalline insulators}

\author{Vassilios Yannopapas}
\email{vyannop@upatras.gr} \affiliation{Department of Materials
Science, University of Patras, GR-26504 Patras, Greece}

\date{\today}

\begin{abstract}
We show that a tetragonal lattice of weakly interacting cavities
with uniaxial electromagnetic response is the photonic counterpart
of topological crystalline insulators, a new topological phase of
atomic band insulators. Namely, the frequency band structure
stemming from the interaction of resonant modes of the individual
cavities exhibits an omnidirectional band gap within which gapless
surface states emerge for finite slabs of the lattice. Due to the
equivalence of a topological crystalline insulator with its
photonic-crystal analog, the frequency band structure of the
latter can be characterized by a $Z_{2}$ topological invariant.
Such a topological photonic crystal can be realized in the
microwave regime as a three-dimensional lattice of dielectric
particles embedded within a continuous network of thin metallic
wires.
\end{abstract}

\pacs{42.70.Qs, 73.20.-r, 73.43.-f} \maketitle
\bibliographystyle{apsrev}

\section{Introduction}

The frequency band structure of artificial periodic dielectrics
formally known as photonic crystals is the electromagnetic (EM)
counterpart of the electronic band structure in ordinary atomic
solids.
Recently, a new analogy between electron and photon states in
periodic structures has been proposed by Raghu and Haldane,
\cite{haldane} namely the one-way chiral edge states in
two-dimensional (2D) photonic-crystal slabs which are similar to
the corresponding edge states in the quantum Hall effect.
\cite{one_way} The photonic chiral edge states are a result of
time-reversal (TR) symmetry breaking which comes about with the
inclusion of gyroelectric/ gyromagnetic material components; these
states are robust to disorder and structural imperfections as long
as the corresponding topological invariant (Chern number in this
case) remains constant.

In certain atomic solids, TR symmetry breaking is not prerequisite
for the appearance of topological electron states as it is the
case in the quantum Hall effect. Namely, when spin-orbit
interactions are included in a TR symmetric graphene sheet, a bulk
excitation gap and spin-filtered edge states emerge
\cite{mele_2005} without the presence of an external magnetic
field, a phenomenon which is known in literature as quantum spin
Hall effect. Its generalization to three-dimensional (3D) atomic
solids lead to a new class of solids, namely, topological
insulators. \cite{ti_papers} The latter possess a
spin-orbit-induced energy gap and gapless surface states
exhibiting insulating behavior in bulk and metallic behavior at
their surfaces.

Apart from topological insulators where the spin-orbit band
structure with TR symmetry defines the topological class of the
corresponding electron states, other topological phases have been
proposed such as topological superconductors (band structure with
particle-hole symmetry), \cite{ts_papers} magnetic insulators
(band structure with magnetic translation symmetry),
\cite{mi_papers} and, very recently, topological crystalline
insulators. \cite{fu_prl} In the latter case the band structure
respects TR symmetry as well as a certain point-group symmetry
leading to bulk energy gap and gapless surface states.

In this work, we propose a photonic analog of a topological
crystalline insulator. Our model photonic system is a 3D crystal
of weakly interacting resonators respecting TR symmetry and the
point-symmetry group associated with a given crystal surface. As a
result, the system possesses an omnidirectional band gap within
which gapless surface states of the EM field are supported. It is
shown that the corresponding photonic band structure is equivalent
to the energy band structure of an atomic topological crystalline
insulator and, as such, the corresponding states are topological
states of the EM field classified by a $Z_{2}$ topological
invariant.

The frequency band structure of photonic crystals whose
(periodically repeated) constituent scattering elements interact
weakly with each other can be calculated by a means which is
similar to the tight-binding method employed for atomic insulators
and semiconductors. Photonic bands amenable to a
tight-binding-like description are e.g., the bands stemming from
the whispering-gallery modes of a lattice of high-index scatterers
\cite{lido} the defect bands of a sublattice of point defects,
within a photonic crystal with an absolute band gap,
\cite{bayindir} the plasmonic bands of a lattice of metallic
spheres \cite{quinten} or of a lattice of dielectric cavities
within a metallic host. \cite{stefanou_ssc}
In the latter case, the frequency band structure stems from the
weak interaction of the surface plasmons of each individual cavity
\cite{stefanou_ssc} wherein light propagates within the crystal
volume by a hopping mechanism. Such type of lattice constitutes
the photonic analog of a topological crystalline insulator
presented in this work whose frequency band structure will be
revealed based on a photonic tight-binding treatment within the
framework of the coupled-dipole method. \cite{cde} The latter is
an exact means of solving Maxwell's equations in the presence of
nonmagnetic scatterers.

\section{Tight-binding description of dielectric cavities in a
plasmonic host}

We consider a lattice of dielectric cavities within a lossless
metallic host. The $i$-th cavity is represented by a dipole of
moment ${\bf P}_{i}=(P_{i;x},P_{i;y},P_{i;z})$ which stems from an
incident electric field ${\bf E}^{inc}$ and the field which is
scattered by all the other cavities of the lattice. This way the
dipole moments of all the cavities are coupled to each other and
to the external field leading to the coupled-dipole equation
\begin{equation}
{\bf P}_{i}= \boldsymbol\alpha_{i}(\omega) [{\bf E}^{inc} +
\sum_{i' \neq i} {\bf G}_{i i'}(\omega) {\bf P}_{i'}].
\label{eq:cde}
\end{equation}
${\bf G}_{i i'}(\omega)$ is the electric part of the free-space
Green's tensor and ${\bf \boldsymbol\alpha}_{i}(\omega)$ is the $3
\times 3$ polarizability tensor of the $i$-th cavity.
Eq.~(\ref{eq:cde}) is a $3N \times 3N$ linear system of equations
where $N$ is the number of cavities of the system. We assume that
the cavities exhibit a uniaxial EM response, i.e., the
corresponding polarizability tensor is diagonal with
$\alpha_{x}=\alpha_{y}=\alpha_{\parallel}$ and
$\alpha_{z}=\alpha_{\perp}$. For strong anisotropy, the cavity
resonances within the $xy$-plane and along the $z$-axis can be
spectrally distinct; thus, around the region of e.g., the cavity
resonance $\omega_{\parallel}$ within the $xy$-plane,
$\alpha_{\perp} \ll \alpha_{\parallel}$ (see appendix). In this
case, one can separate the EM response within the $xy$-plane from
that along the $z$-axis and Eq.~(\ref{eq:cde}) becomes a $2N
\times 2N$ system of equations,
\begin{equation}
{\bf P}_{i}= \alpha_{\parallel}(\omega) [\sum_{i' \neq i} {\bf
G}_{i i'}(\omega) {\bf P}_{i'}]. \label{eq:cde_no_field}
\end{equation}
where we have set ${\bf E}^{inc}={\bf 0}$ since we are seeking the
eigenmodes of the system of cavities. Also, now, ${\bf
P}_{i}=(P_{i;x},P_{i;y})$.

For a particle/cavity of electric permittivity
$\epsilon_{\parallel}$ embedded within a material host of
permittivity $\epsilon_{h}$, the polarizability
$\alpha_{\parallel}$ is given by the Clausius-Mossotti formula
\begin{equation}
\alpha_{\parallel}=\frac{3 V}{4 \pi}
\frac{\epsilon_{\parallel}-\epsilon_{h}}{\epsilon_{\parallel}+
2\epsilon_{h}} \label{eq:cm}
\end{equation}
where $V$ is the volume of the particle/ cavity. For a lossless
plasmonic (metallic) host in which case the electric permittivity
can be taken as Drude-type, i.e., $\epsilon_{h}=1-\omega_{p}^{2} /
\omega^{2}$ (where $\omega_{p}$ is the bulk plasma frequency), the
polarizability $\alpha_{\parallel}$ exhibits a pole at
$\omega_{\parallel}=\omega_{p} \sqrt{2/ (\epsilon_{\parallel}
+2)}$ (surface plasmon resonance). By making a Laurent expansion
of $\alpha_{\parallel}$ around $\omega_{\parallel}$ and keeping
the leading term, we may write
\begin{equation}
\alpha_{\parallel}= \frac{F} {\omega - \omega_{\parallel}} \equiv
\frac{1} {\Omega} \label{eq:a_laurent}
\end{equation}
where $F=(\omega_{\parallel}/2) (\epsilon_{\parallel} -
\epsilon_{h})/ (\epsilon_{\parallel}+2)$. For sufficiently high
value of the permittivity of the dielectric cavity, i.e.,
$\epsilon_{\parallel}
> 10$, the electric field of the surface plasmon is much localized at the surface of the
cavity. As a result, in a periodic lattice of cavities, the
interaction of neighboring surface plasmons is very weak leading
to much narrow frequency bands. By treating such a lattice in a
tight binding-like framework, we may assume that the Green's
tensor ${\bf G}_{i i'}(\omega)$ does not vary much with frequency
and therefore, ${\bf G}_{i i'}(\omega) \simeq {\bf G}_{i
i'}(\omega_{\parallel})$. In this case,
Eq.~(\ref{eq:cde_no_field}) becomes an eigenvalue problem
\begin{equation}
\sum_{i' \neq i} {\bf G}_{i i'}(\omega_{\parallel}) {\bf P}_{i'}=
\Omega {\bf P}_{i} \label{eq:cde_eigen}
\end{equation}
where
\begin{eqnarray}
{\bf G}_{i i'}(\omega_{\parallel})=q_{\parallel}^{3} \Bigl[
C(q_{\parallel} | r_{ii'}|) {\bf
I}_{2} + J(q_{\parallel} | r_{ii'}|) \left(%
\begin{array}{cc}
  \frac{x_{ii'}^2}{r_{ii'}^{2}} & \frac{x_{ii'}y_{ii'}}{r_{ii'}^{2}} \\
  \frac{x_{ii'}y_{ii'}}{r_{ii'}^{2}} & \frac{y_{ii'}^2}{r_{ii'}^{2}} \\
\end{array}%
\right) \Bigr]. \nonumber \\ \label{eq:g_tensor}
\end{eqnarray}
with ${\bf r}_{ii'}={\bf r}_{i}-{\bf r}_{i'}$,
$q_{\parallel}=\sqrt{\epsilon_{h}}\omega_{\parallel}/c$ and ${\bf
I}_{2}$ is the $2 \times 2$ unit matrix. The form of functions
$C(q_{\parallel} | r_{ii'}|)$, $J(q_{\parallel} | r_{ii'}|)$
generally depends on the type of medium hosting the cavities
(isotropic, gyrotropic, bi-anisotropic, etc).
\cite{eroglu,dmitriev} The Green's tensor of
Eq.~(\ref{eq:g_tensor}) describes the electric interactions
between two point dipoles ${\bf P}_{i}$ and ${\bf P}_{i'}$ each of
which corresponds to a single cavity. The first term of ${\bf
G}_{i i'}$ describes an interaction which does not depend on the
orientation of the two dipoles whilst the second one is
orientation dependent.

\small
\begin{figure}[h]
\centerline{\includegraphics[width=8cm]{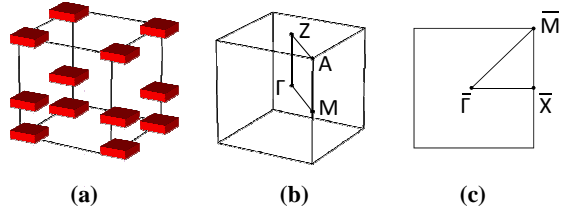}} \caption{(Color
online) (a) Tetragonal crystal with two cavities within the unit
cell. (b) The bulk Brillouin zone and (c) the surface Brillouin
zone corresponding to the (001) surface.} \label{fig1}
\end{figure} \normalsize

For an infinitely periodic system, i.e., a crystal of cavities, we
assume the Bloch ansatz for the polarization field, i.e.,
\begin{equation}
{\bf P}_{i}={\bf P}_{n \beta}=\exp (i {\bf k} \cdot {\bf R}_{n})
{\bf P}_{0 \beta} \label{eq:bloch}
\end{equation}
The cavity index $i$ becomes composite, $i \equiv n \beta$, where
$n$ enumerates the unit cell and $\beta$ the positions of
inequivalent cavities in the unit cell. Also, ${\bf R}_{n}$
denotes the lattice vectors and ${\bf k}=(k_{x},k_{y},k_{z})$ is
the Bloch wavevector. By substituting Eq.~(\ref{eq:bloch}) into
Eq.~(\ref{eq:cde_eigen}) we finally obtain
\begin{equation}
\sum_{\beta'} \tilde{{\bf G}}_{\beta
\beta'}(\omega_{\parallel},{\bf k}) {\bf P}_{0 \beta'}= \Omega
{\bf P}_{0 \beta} \label{eq:cde_eigen_periodic}
\end{equation}
where
\begin{equation}
\tilde{{\bf G}}_{\beta \beta'}(\omega_{\parallel}, {\bf k}) =
\sum_{n'} \exp [i {\bf k} \cdot ({\bf R}_{n}-{\bf R}_{n'})] {\bf
G}_{n \beta; n' \beta'}(\omega_{\parallel}).
\label{eq:green_fourier}
\end{equation}
Solution of Eq.~(\ref{eq:cde_eigen_periodic}) provides the
frequency band structure of a periodic system of cavities.

\small
\begin{figure}[h]
\centerline{\includegraphics[width=8cm]{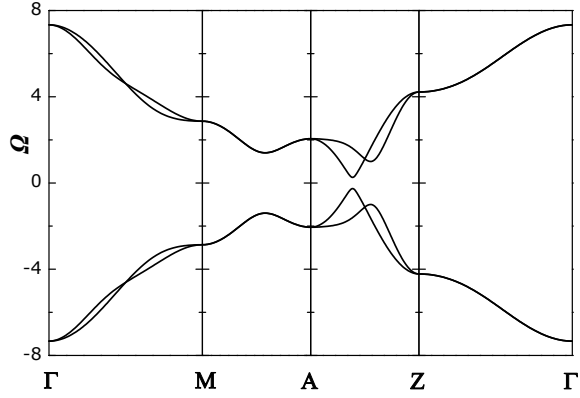}}
\caption{Frequency band structure for tetragonal lattice of
resonant cavities within a plasmonic host (see Fig.~\ref{fig1})
corresponding to the Green's tensor of Eq.~(\ref{eq:G_elem}) with
$s^{A}_{1}=-s^{B}_{2}=1.2, s^{A}_{2}=-s^{B}_{2}=0.5,
s'_{1}=2.5,s'_{2}=0.5,s_{z}=2$.} \label{fig2}
\end{figure}
\normalsize

\section{Topological frequency bands}

Since Eq.~(\ref{eq:cde_eigen_periodic}) is equivalent to a
Hamiltonian eigenvalue problem, we adopt the crystal structure of
Ref.~\onlinecite{fu_prl}. Namely, a tetragonal lattice with a unit
cell consisting of two same cavities at inequivalent positions $A$
and $B$ [see Fig.~\ref{fig1}(a)] along the $c$-axis. In this case,
the index $\beta$ in Eq.~(\ref{eq:cde_eigen_periodic}) assumes the
values $\beta=A,B$ for each sublattice (layer) of the crystal. The
above lattice is characterized by the $C_{4}$ point-symmetry
group. In order to preserve the $C_{4}$ symmetry \cite{fu_prl} in
the Green's tensor matrix of Eq.~(\ref{eq:cde_eigen_periodic}) we
assume that the interaction between two cavities within the same
layer (either $A$ or $B$) depends on the relative orientation of
the point dipole in each cavity whilst the interaction between
cavities belonging to adjacent layers is orientation independent.
Also, we take into account interactions up to second neighbors in
both inter- and intra-layer interactions. Taking the above into
account, the lattice Green's tensor assumes the form
\begin{eqnarray}
\tilde{{\bf G}}({\bf k})
=\left(%
\begin{array}{cc}
  \tilde{{\bf G}}^{AA}({\bf k}) & \tilde{{\bf G}}^{AB}({\bf k}) \\
  \tilde{{\bf G}}^{AB \dagger}({\bf k}) & \tilde{{\bf G}}^{BB}({\bf k}) \\
\end{array}%
\right) \nonumber \\
\end{eqnarray}
where
\begin{eqnarray}
\tilde{{\bf G}}^{\beta \beta}({\bf k})= 2 s^{\beta}_{1}
 \left(
\begin{array}{cc}
  \cos (k_{x} \alpha) & 0 \\
  0 & \cos (k_{y} \alpha) \\
\end{array}
\right)+
\nonumber && \\
2 s^{\beta}_{2}\left(%
\begin{array}{cc}
\cos (k_{x} \alpha)\cos (k_{y} \alpha) & -\sin (k_{x} \alpha)\sin (k_{y} \alpha) \\
  -\sin (k_{x} \alpha)\sin (k_{y} \alpha) & \cos (k_{x} \alpha)\cos (k_{y} \alpha) \\
\end{array}%
\right), 
\nonumber && \\
\tilde{{\bf G}}^{A B}({\bf k})=[s'_{1}+2 s'_{2} (\cos(k_{x}
\alpha) + \cos (k_{y} \alpha)) +s'_{z} \exp( i k_{z} \alpha)] {\bf
I}_{2}. \nonumber && \\ \label{eq:G_elem}
\end{eqnarray}

\small
\begin{figure}[h]
\centerline{\includegraphics[width=8cm]{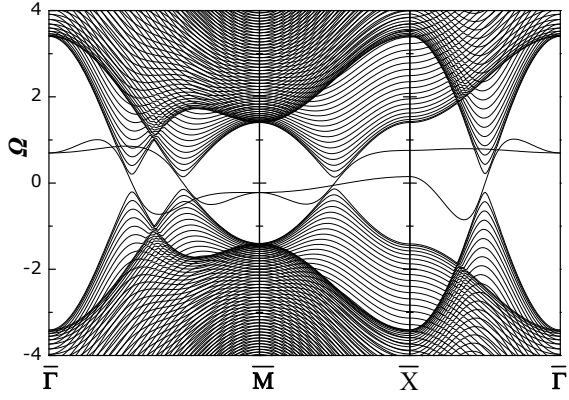}}
\caption{Frequency band structure for a finite slab $ABAB \cdots
ABB$ of the crystal of Fig.~\ref{fig1} made from 80 bilayers. }
\label{fig3}
\end{figure}
\normalsize

The lattice Green's tensor of Eq.~(\ref{eq:G_elem}) is completely
equivalent to the lattice Hamiltonian of Ref.~\onlinecite{fu_prl}.
$s^{\beta}_{1},s^{\beta}_{2},s'_{1},s'_{2},s_{z}$ in
Eq.~(\ref{eq:G_elem}) generally depend on $q_{\parallel}$, the
lattice constant $a$ and the interlayer distance $c$ but hereafter
will be used as independent parameters. Namely we choose
$s^{A}_{1}=-s^{B}_{2}=1.2, s^{A}_{2}=-s^{B}_{2}=0.5,
s'_{1}=2.5,s'_{2}=0.5,s_{z}=2$. In Fig.~\ref{fig2} we show the
(normalized) frequency band structure corresponding to
Eq.~(\ref{eq:G_elem}) along the symmetry lines of the Brillouin
zone shown in Fig.~\ref{fig1}(b). It is evident that an
omnidirectional frequency band gap exists around $\Omega=0$ which
is prerequisite for the emergence of surface states. In order to
inquire the occurrence of surface states we find the eigenvalues
of the Green's tensor of Eq.~(\ref{eq:G_elem}) in a form
appropriate for a slab geometry. The emergence of surface states
depends critically on the surface termination of the finite slab,
i.e., for different slab terminations different surface-state
dispersions occur (if occur at all). Namely, we assume a finite
slab parallel to the (001) surface (characterized by the $C_{4}$
symmetry group) consisting of 80 alternating $AB$ layers except
the last bilayer which is $BB$, i.e., the layer sequence is $ABAB
\cdots ABB$. The corresponding frequency band structure along the
symmetry lines of the surface Brillouin zone of the (001) surface
[see Fig.~\ref{fig1}(c)] is shown in Fig.~\ref{fig3}. It is
evident that there exist gapless surface states within the band
gap exhibiting a quadratic degeneracy at the $\overline{M}$-point.
In this case, the corresponding doublet of surface states can be
described by an effective theory \cite{chong_prb} similarly to the
doublet states at a point of linear degeneracy (Dirac point).
\cite{sepkhanov}

We note that the equivalence of the Green's tensor ${\bf G}$ with
the atomic Hamiltonian of Ref.~\onlinecite{fu_prl} as well as the
form of the time-reversal $T$ and (geometric) $C_{4}$-rotation $U$
operators for the EM problem \cite{haldane} which are the same as
for spinless electrons, allows to describe the photonic band
structure with the $Z_{2}$ topological invariant $\nu_{0}$
\begin{equation}
(-1)^{\nu_{0}}=(-1)^{\nu_{\Gamma M}} (-1)^{\nu_{A Z}}
\label{eq:z2_def}
\end{equation}
where for real eigenvectors of $\tilde{{\bf G}}({\bf k})$ we have
\cite{fu_prl}
\begin{equation}
(-1)^{{\bf k}_{1} {\bf k}_{2}} = {\rm Pf}[w({\bf k}_{2})]/ {\rm
Pf}[w({\bf k}_{1})] \label{eq:pf_frac}
\end{equation}
and
\begin{equation}
w_{mn}({\bf k}_{i}) = \langle u_{m} ({\bf k}_{i}) | U | u_{n}({\bf
k}_{i}) \rangle. \label{eq:w_def}
\end{equation}
${\rm Pf}$ stands for the Pfaffian of a skew-symmetric matrix,
i.e., ${\rm Pf}[w]^{2}=\det(w)$. Due to the double degeneracy of
the band structure at the four special momenta points $\Gamma, M,
A, Z$ the frequency bands come in doublets. Since frequency
eigenvectors with different eigenfrequencies are orthogonal, all
the inter-pair elements of the $w$-matrix are zero and the latter
is written as:
\begin{equation}
w({\bf k}_{i})=\left(%
\begin{array}{cccc}
 w^{1}({\bf k}_{i})  & 0 & 0 & 0 \\
 0  & w^{2}({\bf k}_{i}) & 0 & 0 \\
 0  & 0 & \ddots & 0 \\
  0 & 0 & 0 & w^{N}({\bf k}_{i}) \\
\end{array}%
\right) \label{eq:w_reduced_form}
\end{equation}
where $w^{j}({\bf k}_{i})$ are anti-symmetric $SU(2)$ matrices,
\cite{wang_njp} i.e., $w^{j}({\bf k}_{i})=A_1$ or $A_2$, where
\begin{equation}
A_1= \left(%
\begin{array}{cc}
  0 & 1 \\
  -1 & 0 \\
\end{array}%
\right), A_2= \left(%
\begin{array}{cc}
  0 & -1 \\
  1 & 0 \\
\end{array}%
\right). \label{eq:alpha_matr_def}
\end{equation}
In this case, ${\rm Pf}[w({\bf k}_{i})]=w^{1}_{12} w^{2}_{12}
\cdots w^{N}_{12}=\pm 1$. Therefore, $(-1)^{{\bf k}_{1} {\bf
k}_{2}} = \pm 1$ and $\nu_{0}=1$ which ensures the presence of
gapless surface states.

We note that the above analysis relies on the assumption of real
frequency bands. The presence of losses in the constituent
materials renders the frequency bands complex, i.e., the Bloch
wavevector possesses both a real and an imaginary part. However,
even in this case, one can still speak of real frequency bands if
the imaginary part of the Bloch wavevector is at least {\it
hundred} times smaller than the corresponding real part. This a
common criterion used in calculations of the complex frequency
band structure by on-shell electromagnetic solvers such as the
layer-multiple scattering method \cite{comphy} or the
transfer-matrix method. \cite{tmm}

\small
\begin{figure}[h]
\centerline{\includegraphics[width=6cm]{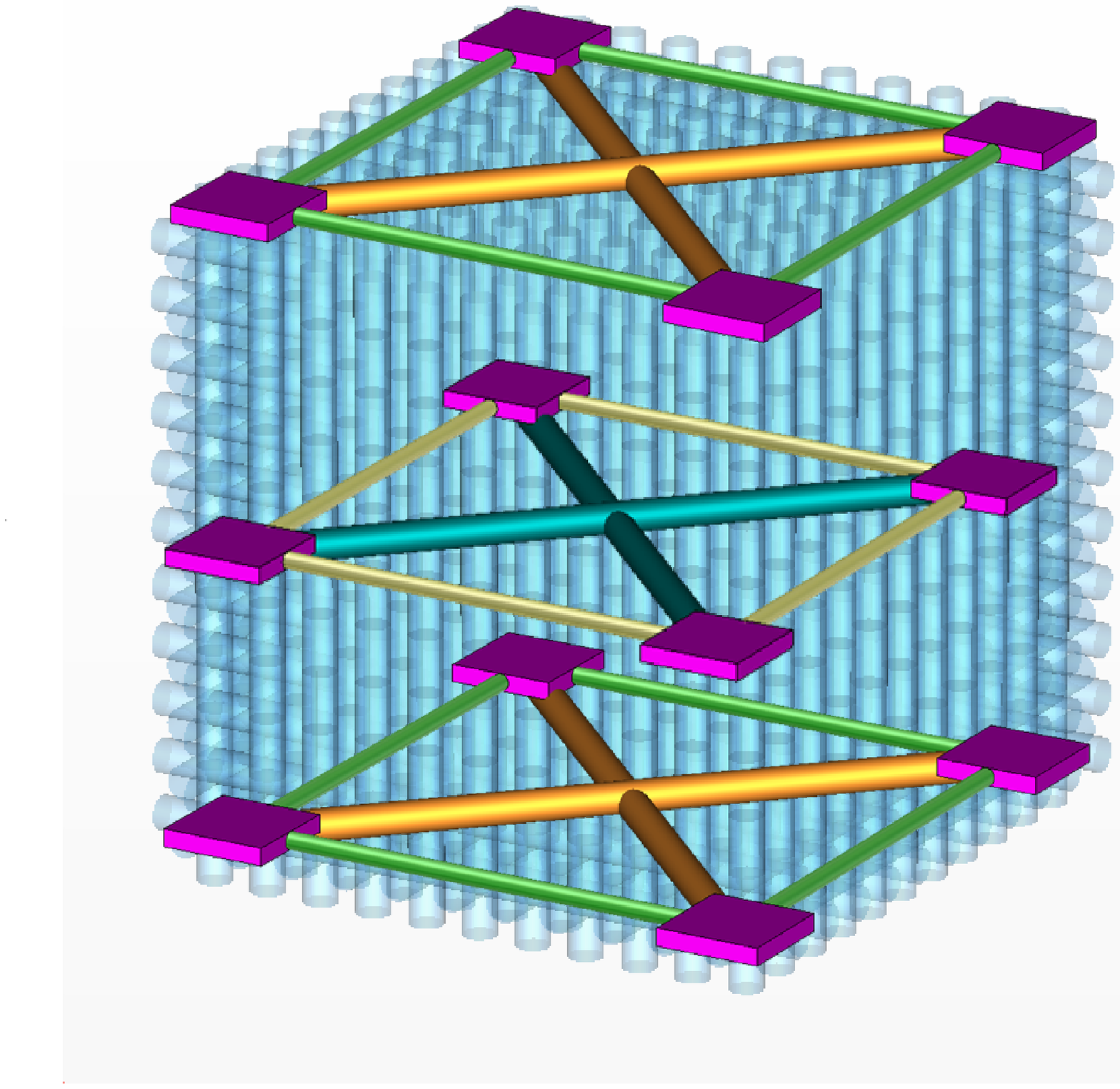}} \caption{(Color
online) A possible realization of a photonic structure with
gapless surface states: dielectric particles of square cross
section, joined together with cylindrical coupling elements and
embedded within a 3D network of metallic wires (artificial
plasma).} \label{fig4}
\end{figure}
\normalsize

\section{Blueprint for a photonic topological insulator}

A possible realization of the photonic analogue of topological
insulator in the laboratory is depicted in Fig.~\ref{fig4}. Since
our model system requires dielectric cavities within a homogeneous
plasma, a lattice of nano-cavities formed within a homogeneous
Drude-type metal, e.g., a noble metal (Au, Ag, Cu), would be the
obvious answer. \cite{stefanou_ssc} However, the plasmon bands are
extremely lossy due to the intrinsic absorption of noble metals in
the visible regime. A solution to this would be the use of an {\it
artificial} plasmonic medium operating in the microwave regime
where metals are perfect conductors and losses are minimal.
Artificial plasma can be created by a 3D network of thin metallic
wires of a few tens of $\mu$m in diameter and spaced by a few mm.
\cite{art_plasma} A lattice of dielectric particles within an
artificial plasma can be modelled with the presented tight-binding
Green's tensor. Since the interaction among first and second
neighbors within the same bilayer ($A$ or $B$) should depend on
the dipole orientation (in order to preserve the $C_{4}$
symmetry), the dielectric particles in each layer are connected
with cylindrical waveguiding elements (different in each layer $A$
or $B$ - see Fig.~\ref{fig4}). In contrast, between two successive
bilayers there are no such elements since interactions between
dipoles belonging to different layers should be independent of the
dipole orientations (s orbital-like). Another advantage of
realizing the photonic analog in the microwave regime is the
absence of nonlinearities in the EM response of the constituent
materials since photon-photon interactions may destroy the
quadratic degeneracy of the surface bands in analogy with
fermionic systems. \cite{sun}

Finally, we must stress that a photonic topological crystalline
insulator can be also realized with purely dielectric materials if
the host medium surrounding the cavities is not a plasmonic medium
but a photonic crystal with an absolute band gap: the cavities
would be point defects within the otherwise periodic photonic
crystal and the tight-binding description would be still
appropriate. \cite{bayindir,k_fang} In this case, Maxwell's
equations lack of any kind of characteristic length and the
proposed analog would be realized in any length scale.

\section{Conclusions}

In conclusion, a 3D lattice of weakly interacting cavities
respecting TR symmetry and a certain point-group symmetry
constitutes a photonic analog of a topological crystalline
insulator by demonstrating a spectrum of gapless surface states. A
possible experimental realization would be a 3D lattice of
dielectric particles within a continuous network of thin metallic
wires with a plasma frequency in the GHz regime.

This work has been supported by the European Community's Seventh
Framework Programme (FP7/2007-2013) under Grant Agreement No.
228455-NANOGOLD (Self-organized nanomaterials for tailored optical
and electrical properties).

\appendix*
\section{}
The $z$-component of the polarizability need not be zero but can
assume finite values as far as the frequency band stemming from
the surface-plasmon resonance corresponding to the z-component is
spectrally distinct (no overlap) with the bands stemming from the
$xy$-components of the polarizability. In this case, one can treat
separately the two frequency bands (doublet) stemming from the
xy-resonance from the band coming from the z-resonance (singlet).
The above requirements can be quantified as follows,
\begin{equation}
C_{\parallel} \ll \frac{\omega_{\parallel} -
\omega_{\perp}}{\omega_{p}} \label{eq:app_1}
\end{equation}
where $C_{\parallel}$  is the width of the $xy$-frequency bands
(in dimensionless frequency units). $C_{\parallel}$ is obtained
from the first term of the right-hand side of
Eq.~(\ref{eq:g_tensor}) of the paper. To a first approximation, it
is given by \cite{cde}
\begin{equation}
C_{\parallel} \sim \frac{\exp(-q_{\parallel} a_{\parallel})}
{q_{\parallel} a_{\parallel}} \label{eq:app_2}
\end{equation}
where $q_{\parallel}= \sqrt{|\epsilon_{h}|} \omega_{\parallel} /
c$. Therefore, the condition (\ref{eq:app_1}) is written as
\begin{equation}
\frac{\exp(-q_{\parallel} a_{\parallel})} {q_{\parallel}
a_{\parallel}} \ll \frac{\omega_{\parallel} -
\omega_{\perp}}{\omega_{p}} \label{eq:app_3}
\end{equation}
where $a_{\parallel}$ is the lattice constant in the $xy$-plane.
Given that $\omega_{\parallel} = \omega_{p}
\sqrt{2/(\epsilon_{\parallel}+2)}$, Eq.~(\ref{eq:app_3}) becomes
\begin{equation}
\frac{\exp({-\sqrt{|\epsilon_{h}|} \omega_{\parallel}
a_{\parallel} /c})}{\sqrt{|\epsilon_{h}|} \omega_{\parallel}
a_{\parallel} /c} \ll \sqrt{\frac{2
(\epsilon_{\perp}-\epsilon_{\parallel})}
{(\epsilon_{\parallel}+2)(\epsilon_{\perp}+2)}} \label{eq:app_5}
\end{equation}
From the above equation it is evident that for a given value of
the dielectric anisotropy $\epsilon_{\perp} -
\epsilon_{\parallel}$, one can always find a suitably large
lattice constant $\alpha_{\parallel}$ such that
Eq.~(\ref{eq:app_5}) is fulfilled. The latter allows the easy
engineering of the photonic analog of a crystalline topological
insulator since there is practically no restriction on the choice
of the (uniaxial) material the cavities are made from. It can also
be easily understood that if Eq.~(\ref{eq:g_tensor}) holds, the
same equation is true for the width $C_{\perp}$  of the singlet
frequency band (resulting from the $z$-resonance).

{\it Numerical example}. Suppose that the cavities are made from a
nematic liquid crystal which is a uniaxial material. Typical
values of the permittivity tensor $\epsilon$ are e.g.,
$\epsilon_{\parallel}=1.5$, $\epsilon_{\perp}=1.8$. In this case,
$\omega_{\parallel} \approx 0.75 \omega_{p}$ and
$\epsilon_{h}=1-\omega_{p}^{2} / \omega_{\parallel}^{2} \approx
-0.777$. By choosing a large lattice constant, i.e.,
$a_{\parallel} = 4 c / \omega_{p}$ , Eq.~(\ref{eq:app_5}) is
clearly fulfilled
\begin{equation}
\frac{\exp{(-\sqrt{|\epsilon_{h}|} \omega_{\parallel}
a_{\parallel} /c)}}{\sqrt{|\epsilon_{h}|} \omega_{\parallel}
a_{\parallel} /c} \approx 0.026866 \ll 0.2213 \approx
\sqrt{\frac{2 (\epsilon_{\perp}-\epsilon_{\parallel})}
{(\epsilon_{\parallel}+2)(\epsilon_{\perp}+2)}}. \label{eq:app_6}
\end{equation}


\begin{thebibliography}{}
\bibitem{haldane} F.~D.~M.~Haldane and S.~Raghu, \prl {\bf 100},
013904 (2008); {\it ibid}, \pra {\bf 78}, 033834 (2008).
\bibitem{one_way} Z.~Wang, Y.~D.~Chong, J.~D.~Joannopoulos, and M.~Solja\v{c}i\'{c}, \prl
{\bf 100}, 013905 (2008); {\it ibid}, Nature (London) {\bf 461},
772 (2009);
Z.~Yu {\it et al.}, \prl {\bf 100}, 023902 (2008); H.~Takeda and
S.~John, \pra {\bf 78}, 023804 (2008); D.~Han {\it et al.}, \prl
{\bf 102}, 123904 (2009); X.~Ao, Z.~Lin, and C.~T.~Chan, \prb {\bf
80}, 033105 (2009); M.~Onoda and T.~Ochiai, \prl {\bf 103}, 033903
(2009); T.~Ochiai and M.~Onoda, \prb {\bf 80}, 155103 (2009);
R.~Shen, L.~B.~Shao, B.~Wang, and D.~Y.~Xing, \prb {\bf 81},
041410(R) (2010);
Y.~Poo {\it et al.}, \prl {\bf 106}, 093903 (2011).
\bibitem{mele_2005} C.~L.~Kane and E.~J.~Mele, \prl {\bf 95},
226801 (2005); {\it ibid}, \prl {\bf 95}, 146802 (2005).
\bibitem{ti_papers} L.~Fu, C.~L.~Kane, and E.~J.~Mele, \prl {\bf
98}, 106803 (2007); L.~Fu and C.~L.~Kane, \prb {\bf 76}, 045302
(2007); M.~Z.~Hasan and C.~L.~Kane, \rmp {\bf 82}, 3045 (2010).
\bibitem{ts_papers} A.~P.~Schnyder, S.~Ryu, A.~Furusaki, and
A.~W.~W.~Ludwig, \prb {\bf 78}, 195125 (2008); A.~Kitaev,
AIP~Conf.~Proc. {\bf 1134}, 22 (2009); Y.~Ran, arXiv:~1006.5454;
X.-~L.~Qi, T.~L.~Hughes, S.~Raghu, and S.-~C.~Zhang, \prl {\bf
102}, 187001 (2009); R.~Roy, arXiv:~0803.2868.
\bibitem{mi_papers} R.~S.~K.~Mong, A.~M.~Essin, and J.~E.~Moore,
\prb {\bf 81}, 245209 (2010); ; R.~Li, J.~Wang, X.-~L.~Qi, and
S.-~C.~Zhang, Nat.~Phys. {\bf 6}, 284 (2010).
\bibitem{fu_prl} L.~Fu, \prl {\bf 106}, 106802 (2011).
\bibitem{lido} E.~Lidorikis, M.~M.~Sigalas, E.~N.~Economou, and
C.~M.~Soukoulis, \prl {\bf 81}, 1405 (1998).
\bibitem{bayindir} M.~Bayindir, B.~Temelkuran, and E.~Ozbay, \prl
{\bf 84}, 2140 (2000).
\bibitem{quinten} M.~Quinten, A.~Leitner, J.~R.~Krenn,
and F.~R.~Aussenegg, Opt.~Lett.~{\bf 23}, 1331 (1998).
\bibitem{stefanou_ssc} N.~Stefanou, A.~Modinos, and V.~Yannopapas,
Sol.~Stat.~Commun. {\bf 118}, 69 (2001).
\bibitem{cde} E.~M.~Purcell and C.~R.~Pennypacker, Astrophys.~J.
{\bf 186}, 705 (1973).
\bibitem{suppl} For a detailed study see supplemental material at
...
\bibitem{eroglu} A.~Eroglu, {\em Wave Propagation and Radiation in Gyrotropic and Anisotropic Media\/}
(Springer, New York, 2010).
\bibitem{dmitriev} V.~Dmitriev, Prog.~Electromag.~Res. {\bf 48},
145 (2004).
\bibitem{chong_prb} Y.~D.~Chong, X.-~G.~Wen, and M.~Soljacic, \prb
{\bf 77}, 235125 (2008).
\bibitem{sepkhanov} R.~A.~Sepkhanov, Ya.~B.~Bazaliy, and
C.~W.~J.~Beenakker, \pra {\bf 75}, 063813 (2007).
\bibitem{wang_njp} Z.~Wang, X-.~L.-~Qi, and S.-~C.-~Zhang,
New~J.~Phys. {\bf 12}, 065007 (2010).
\bibitem{comphy} N.~Stefanou, V.~Yannopapas, and A.~Modinos,
Comput.~Phys.~Commun. {\bf 113}, 49 (1998).
\bibitem{tmm} P.~M.~Bell {\it et al.}, Comput.~Phys.~Commun. {\bf 85}, 306
(1995).
\bibitem{art_plasma} J.~B.~Pendry, A.~J.~Holden, W.~J.~Stewart,
and I.~Youngs, \prl {\bf 76}, 4773 (1996).
\bibitem{sun} K.~Sun, H.~Yao, E.~Fradkin, and S.~A.~Kivelson, \prl
{\bf 103}, 046811 (2009).
\bibitem{k_fang} K.~Fang, Z.~Yu, and S.~Fan, \prb {\bf 84}, 075477
(2011).
\end{thebibliography}
\end{document}